\documentstyle[11pt,newpasp,twoside,psfig]{article}
\def\edcomment#1{\iffalse\marginpar{\raggedright\sl#1\/}\else\relax\fi}
\reversemarginpar

\begin{document}
\title{Coherent Mechanisms of Pulsar Radio Emission}
\author{Maxim Lyutikov$^{1,2}$ }
\author{Roger Blandford $^1$}
\author{George Machabeli $^3$}
\affil{1. CITA
60 St. George street, Toronto, Canada}
\affil{2. Caltech,
Pasadena, California 91125}
\affil{3. Abastumani Astrophysics Observatory, A. Kazbegi Av. 2a, Tbilisi,
380060
 Republic of Georgia}

\begin{abstract}

Relativistic plasma masers operating 
 on the anomalous cyclotron-Cherenkov resonance
$\, \omega-\, k_{\parallel} v_{\parallel} +\, \omega_B/\, \gamma_{res}=0$
and the  Cherenkov-drift resonance
$\, \omega-\, k_{\parallel} v_{\parallel} - k_x u_d =0$,
 are capable of explaining the main observational
characteristics of pulsar radio emission.
 Both  electromagnetic
instabilities are due to the interaction of the fast particles from
the primary beam and from the tail of the secondary pairs distribution 
 with the  normal modes of a strongly magnetized one-dimensional
electron-positron plasma. 
In a typical pulsar both resonances 
 occur in the outer parts of
magnetosphere at $ r_{res}\, \approx 10^9 {\rm cm}$. 

\end{abstract}

\section{ Emission mechanisms}

We have shown  (Lyutikov, Blandford \& Machabeli 1999b)
that 
pulsar radiation may be
generated  by two kinds of
{\it electromagnetic}  plasma instabilities --
cyclotron-Cherenkov instability, developing at the 
anomalous Doppler resonance 
\begin{equation}
\omega({\bf k})- k _{\parallel} V_{\parallel} + \omega  _{B}/\gamma_{res}=0
\label{gr}
\end{equation}
 and Cherenkov-drift instability,
developing at the 
Cherenkov-drift resonance
\begin{equation}
\omega({\bf k})- k _{\parallel} V_{\parallel} -k_x u_d=0
\label{gr1}
\end{equation}
 The
cyclotron-Cherenkov instability is responsible for the
generation of the core-type emission and the
Cherenkov-drift instability is responsible for the
generation of the cone-type emission.
These {\it electromagnetic}
instabilities
are  the strongest instabilities
in the pulsar magnetosphere (Lyutikov 1999a).

Both 
instabilities are maser-type
in   a sense that an induced emission dominates over spontaneous. 
From a classical viewpoint, a random  incoming wave forces a particle to emit
a wave "in phase" with initial one, so that the resulting intensity is 
proportional to $N^2$ where $N$ is a number of the 
interfering  waves. This is how
coherence of the radiation is produced: masers naturally
produce coherent waves. For  the operation of  a maser 
 some kind of  population inversion condition should be satisfied:
there should be more emitting that absorbing particles. 

The cyclotron instability
 can develop on both primary beam and on the tail of the plasma distribution while 
the Cherenkov-drift instability develops on the rising part of the primary beam
distribution function.
The free energy for the growth of the instability
comes from the nonequilibrium
anisotropic distribution of fast particles. 

From the microphysical point of view
 both emission  process habe  more similarities with Cherenkov-type  emission
than with synchrotron or curvature emission. 
 In the case of   Cherenkov-type
process  the emission may be  attributed to the electromagnetic 
polarization shock front that 
develops in a dielectric medium due  to the passage of a charged particle with  
speed larger than phase speed of  waves in a medium. It is  a collective
emission process in which all particles of plasma take part. 

Interestingly, 
in a cyclotron-Cherenkov emission process
an emitting particle undergoes a  transition {\it up} in Landau  levels, thus, 
population inversion condition in this case requires more particles on the {\it lower} levels -
this condition is satisfied by the one dimensional distribution
of particles in pulsar magnetosphere. The  cyclotron-Cherenkov emission  is not new in astrophysics:
it is exactly by this resonance that cosmic rays produce Alfv\'{e}n wave in the interstellar shocks.
In addition, the laboratory devices called 
Slow Wave Electron Cyclotron Masers use this resonance to produce high power microwave
emission.

Emission of a charged particle  propagating {\it in a medium with  a 
 curved magnetic  
field } differs from  conventional Cherenkov, cyclotron or curvature
emission and 
includes, to some extent, the features of each of these mechanisms. 
We have developed a formalism for considering an emissivity of a particle in a curved
field in a medium (Lyutikov, Machabeli \& Blandford  1999b).
 The resulting process may be called a coherent curvature emission.
The Cherenkov-drift  instability that operates  in pulsars is related to some low
frequency electromagnetic instabilities in TOKAMAKs; the ultrarelativistic energies
in pulsars change it to a high frequency one.

Both 
 instabilities develop in a limited
region on the open field lines at large distances from the surface $r \approx 10^9$ cm.
 The size of the emission
region is determined by the curvature of the magnetic
field lines, which limits the length of the resonant wave-particle
interaction and growth of the wave. The location of the cyclotron-Cherenkov 
instability is limited to the field lines with large curvature,
while the Cherenkov-drift instability occurs on the field lines
with the radius of  curvature 
limited both from above and from below. There are two possible locations
of the Cherenkov-drift instability: in a ringlike region around the 
straight field lines and in the region of swept back field lines (Fig.  1).
Thus, both instabilities produce narrow pulses, though they 
operate at radii where the opening angle of the open field  lines
is large.

The proposed theory is capable of explaining many observational results:
\vspace{-.3cm}

\begin{itemize}
 \item 
Energetic: approximately $1\%$ of the beam energy is transfered into radiation
\vspace{-.3cm} \item 
Waves are emitted approximately half way through the light cylinder;
curvature of filed lines controls
the emission regions (cyclotron absorption may also be important)
\vspace{-.3cm} \item Radius-to-frequency mapping for core $  r \propto \nu ^{-1/6}$
\vspace{-.3cm} \item Width-frequency dependence is controlled by the combination of the growth rate of the instability and
the structure of the field line
\vskip -.8 truein
\begin{figure}[h]
\hskip 1 truein
\psfig{file=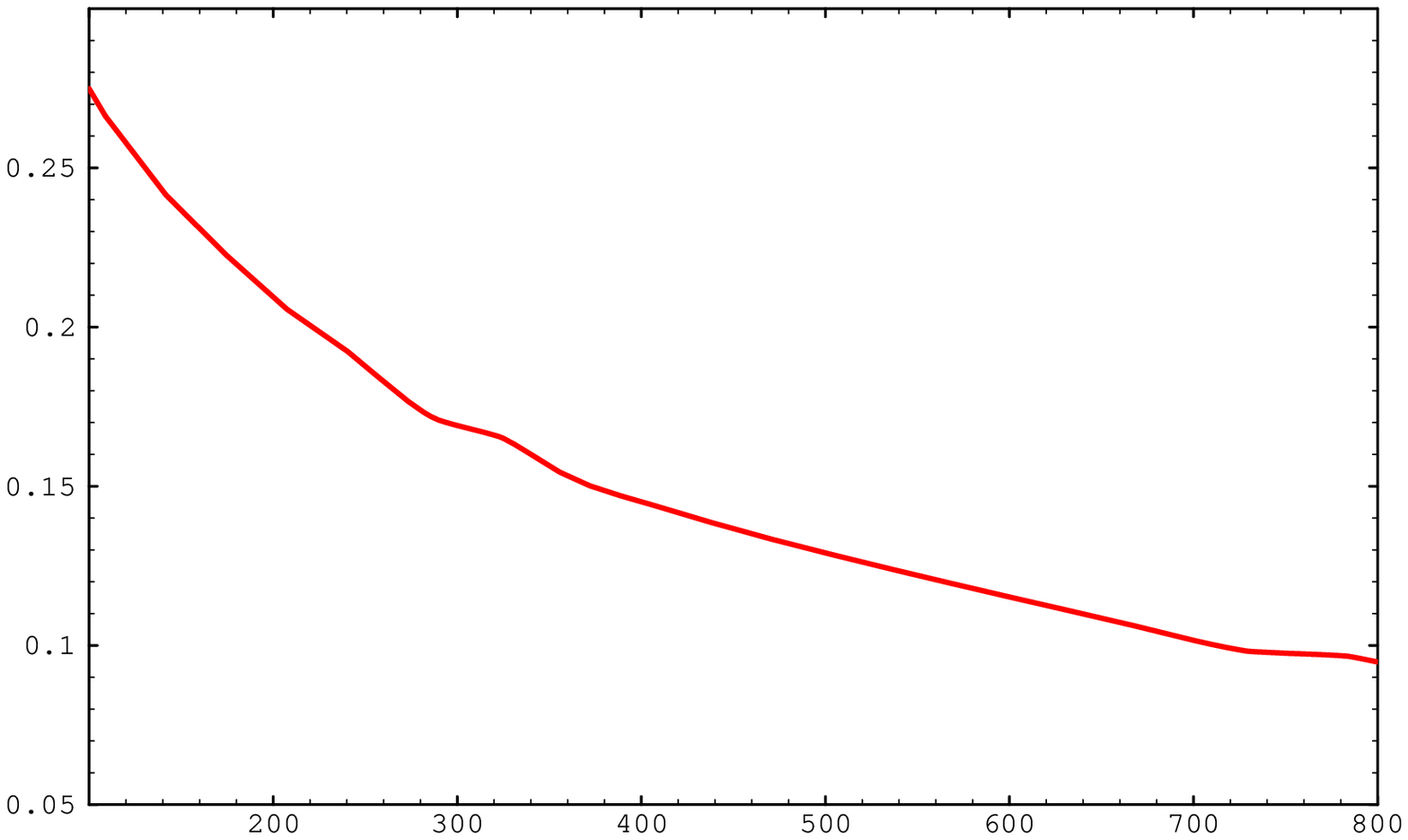,width=5cm}
\end{figure} \vskip -0.8 truein

\vskip -0.8 truein
{\small W, {\rm rad}}
\vskip .4 truein

\hskip 2  truein
{\small MHz}

\vspace{-.3cm} \item Fundamental modes are linearly polarized 
\vspace{-.3cm} \item Different distribution function of secondary electrons and positrons results in circular 
polarization for some $\theta\,<\, \theta^*$. For the resonance on the beam the circular
polarization reaches maximum in the center. For the resonance on the tail particles
the sense of the polarization changes due to the curvature drift.
\vspace{-.3cm} \item Higher  degree of   circular polarization at high frequencies: higher frequencies
resonate only with the beam, lower frequencies can resonate with the tail particles where
both types of charges are present.
\vspace{-.3cm} \item Spectra of the core emission are controled by the quasilinear diffusion, which gives
a spectral index $\alpha=-2$  (Lyutikov 1998a) . Other nonlinear processes, like Raman scattering
 (Lyutikov 1998b) may also be important.
\vspace{-.3cm} \item Large emitting size and high altitudes (Gwinn et al. 1997, Smirnova et al. 1995, Kijak \& Gil 1998)
\vspace{-.3cm} \item High energy emission: development of the cyclotron instability excites  gyration; 
reemission on the normal Dopper resonance boosts the frequency into the optical-UV-soft X-ray
region $\omega \sim \omega_B \gamma$.
\end{itemize}

One of the most challenging consequences of our model is that emission is generated at 
high altitudes. Below we list both observational and theoretical arguments  that support
this: 
\begin{itemize}
\vspace{-.3cm} \item {"Wide beam"} geometry:
 correlation in intensity and position angle in 
 widely separated pulses (Manchester 1995)
\vspace{-.3cm} \item  Emission bridge between some widely separated pulses 
\vspace{-.3cm} \item  {Extra peaks in Crab} at high frequencies 
\vspace{-.3cm} \item  {Alignment} of Radio and High Energy emission in {Crab} and partially in Vela
\vspace{-.3cm} \item Large {emission size} of Vela {(500 km)} (Gwinn et al.  1997) (also 
 Smirnova et al. 1995, Kijak \& Gil 1998)
\vspace{-.3cm} \item For small $r$ typical plasma frequencies $\omega_p\,, \omega_B \gg \omega$
(Kunzl et al. 1998); excitaion of wave with $\omega \ll \omega_p$
is impossible (Lyutikov 1999b)
\end{itemize}

Predictions of our  model are:
\begin{itemize}
\vspace{-.3cm} \item "frequency incoherent maser": at each moment emission consists of narrow frequency features
\vspace{-.3cm} \item increase of circular polarization with frequency (at higher frequencies the cyclotron instability
on the beam with one sign of charge dominates over cyclotron instability on the tail particles, where
both signs of charge present)
\vspace{-.3cm} \item linear polarization of cone  $\perp$ to {\bf B}  plane 
(may be resolved using interstellar scintillations similar to  Smirnova et al. 1996)
\end{itemize}

The
weak points of the model are: (i) 
the development of the cyclotron instability requires that the 
plasma be relatively dense and slow streaming ($\gamma_p \approx 10$) - this is unusual
but not unreasonable;
(ii)  width-period dependence - now it is determined not only by the geometrical factors, but
also by the plasma parameter (growth rates).

\vskip -.6 truein
\begin{figure}[h]
\psfig{file=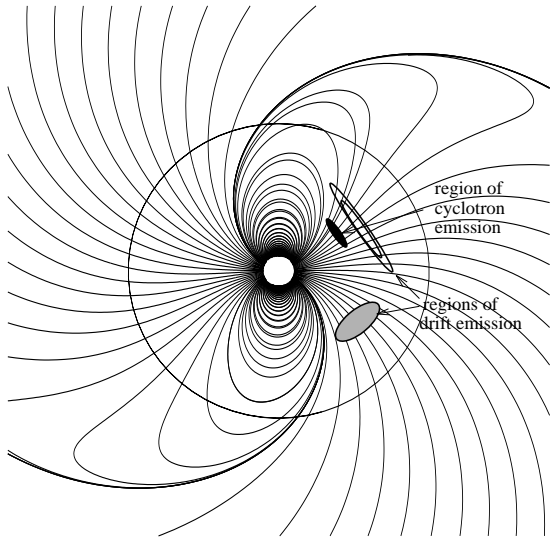,width=7cm}
\caption{Location of the emitting regions}
\label{Deu}
\end{figure}

\end{document}